\documentclass[aps,pre,floatfix]{revtex4}
\usepackage{graphicx}

\usepackage{dcolumn}
\usepackage{bm}
\usepackage{amssymb}
\usepackage{amsmath}
\def\LM#1#2{\left|\begin{array}{l}{#1}\\[1ex]{#2}\end{array}\right.}

\begin{document}
\title{Properties of the reaction
front in a reaction-subdiffusion process
}

\author{Katja Lindenberg$^{1}$ and S. B. Yuste$^{2}$}
\affiliation{${(1)}$
Department of Chemistry and Biochemistry, and Institute for
Nonlinear Science,
University of California San Diego, 9500 Gilman Drive, La Jolla, CA
92093-0340, USA\\
$^{(2)}$
Departamento de F\'{\i}sica, Universidad de Extremadura,
E-06071 Badajoz, Spain
}

\begin{abstract}
We study the reaction front for the process $A+B\to C$
in which the reagents move subdiffusively.  We propose
a fractional reaction-subdiffusion equation in which both
the motion and the reaction terms are affected by the subdiffusive
character of the process.  Scaling solutions to these
equations are presented and compared with those of a direct numerical
integration of the equations.  We find that for reactants whose
mean square displacement varies sublinearly with time as $\langle
r^2\rangle \sim t^\gamma$, the scaling behaviors of the reaction front
can be recovered from those of the corresponding
diffusive problem with the substitution $t\to t^\gamma$.

\end{abstract}

\maketitle


\section{Introduction}
\label{introduction}
It is very well known that
diffusion-limited binary reactions in low dimensions
exhibit ``anomalous'' rate laws, that is, rate laws that deviate from
the quadratic forms that usually characterize binary reactions.  It is
also known that these anomalies occur because diffusion is not an
effective mixing mechanism, especially in constrained geometries,
whereas the quadratic rate laws assume
perfect mixing.  The shortcomings of diffusion as a mixing process lead
to the spontaneous occurrence of spatial order and spatial
structures in terms of which the anomalous rate laws are well
understood~\cite{kotomin}.  Broadly speaking, spatial order
leads to a slowing down of the reaction because it tends to decrease
or deplete the interfacial areas where reactions can occur.

The anomalies exhibit themselves in different ways according to the
design of the experiment, and the critical dimension above which the
behavior of the reaction is ``normal'' also varies from one experiment
to another.  Thus, for example, in an irreversible batch reaction
$A+B\to C$, the decay of reactant concentrations deviates from (is
slower than) the mean field behavior $\rho\sim t^{-1}$ up to dimension
$d=4$. This anomaly is accompanied by and directly associated with
a sharp segregation of species as the
majority species in any region eliminates the minority species and
diffusion is too slow to replenish the minority population effectively.
If there are random sources of reactants so that the system achieves a
steady state, species segregation is observed in the steady state up to
dimension $d=2$~\cite{kotomin,we1}.

One of the more accessible experimental setups to
observe these anomalies is that of a reaction front in which initially
one reactant uniformly occupies all the space to one side of a sharp
front, and the other reactant occupies the other~\cite{galfi}. As the reaction
proceeds, one can monitor the concentration profiles of both reactants
and of the product. Their evolution reflect the anomalies.  For
the reaction-diffusion case the front evolution has been extensively studied
theoretically and confirmed experimentally~\cite{we2}.

Our interest lies in understanding binary reaction kinetics in
media where the reactants move \emph{subdiffusively}.  Subdiffusive
motion is particularly important in the context of complex
systems such as glassy and disordered materials, in which pathways
are constrained for geometric or for energetic reasons.  It is also
particularly germane to the way in which
experiments in low dimensions have to
be carried out.  Such experiments must avoid any active or
convective or advective mixing so as to ensure that any
mixing is only a consequence of diffusion.  To accomplish this
usually requires the use of gel substrates and/or highly constrained
geometries (the first gel-free experiments were carried out
recently~\cite{gelfree,baroud}).  Under
these circumstances it is not clear whether the motion of the species is
actually diffusive, or if it is in fact \emph{subdiffusive}.  Indeed, a
recent detailed discussion on ways to extract accurate parameters
and exponents from such experiments concludes that at least
the experiments presented in that work, carried out in a gel,
reflect subdiffusive rather than diffusive motion~\cite{recent}.

Subdiffusive motion is characterized by a mean square displacement
that varies sublinearly with time,
\begin{equation}
\langle r^2(t)\rangle \sim \frac{2K_\gamma}{\Gamma(1+\gamma)} t^\gamma ,
\label{meansquaredispl}
\end{equation}
with $0<\gamma<1$.  For ordinary diffusion $\gamma=1$, and $K_1\equiv D$
is the ordinary diffusion coefficient.
Subdiffusion is not modeled in a universal way in the literature.
Among the more successful approaches to the subdiffusion problem
have been continuous time random walks with non-Poissonian
waiting time distributions~\cite{Kehr,Bouchaud,Barkai,Sokolov}, and fractional
dynamics approaches in which the ordinary diffusion operator is replaced by
a generalized \emph{fractional} diffusion
operator~\cite{Barkai,Metzler,West,Hilfer,Sch}.
The relation between the two has also been discussed.
In particular, the fractional dynamics formulation that leads to the
mean square displacement~(\ref{meansquaredispl}) can be associated with
a continuous time random walk with a waiting time distribution between
steps which at long times behaves as
\begin{equation}
\psi(t)\sim t^{-\gamma-1}.
\label{waitingtime}
\end{equation}
In our work we adopt the fractional dynamics approach.

Since subdiffusion is an even poorer mixing mechanism than ordinary
diffusion, we expect that the diffusive anomalies observed in constrained
geometries will in some sense be exacerbated in a subdiffusive
environment. However, the outcome is not entirely clear or predictable. On
the one hand, slower motion naturally leads to a slowing down of the
reaction. On the other, slower motion also delays the elimination of the
minority species in a region, and therefore may reduce the spatial
segregation effects observed in the diffusive case.

In this paper we focus on the time dependence of the front geometry,
and, in particular, on the exponents that describe the evolution of a
number of quantities that define the reaction front.  We expect there to
be anomalies that differ from those of the diffusive problem.  In
Sec.~\ref{evolution} we formulate our model and present our scaling
solutions. In Sec.~\ref{comparisons} we compare our scaling results with
those obtained by direct numerical integration of our model equations.
We end with some comments on our results in Sec.~\ref{endnotes}

\section{Evolution of subdiffusive front}
\label{evolution}

We start with a system of $A$ particles on one side and $B$ particles on
the other of a sharp linear front, defined to lie
perpendicular to the $x$ axis.  The particles diffuse and react
with a given probability upon encounter.  A standard mean-field model
for the
evolution of the concentrations $a(x, t)$ and $b(x, t)$
of $A$ and $B$ particles along $x$ is given by the reaction-diffusion
equations
\begin{equation}
\label{difmft}
\begin{array}{rcl}
\displaystyle\frac{\partial}{\partial t} a(x,t)&=& D
\displaystyle\frac{\partial^2}{\partial x^2} a(x,t)-k a(x,t) b(x,t) \\
\noalign{\smallskip}
\displaystyle\frac{\partial}{\partial t} b(x,t)&=& D
\displaystyle\frac{\partial^2}{\partial x^2} b(x,t)-k a(x,t) b(x,t) \; ,
\end{array}
\end{equation}
where $D$ is the diffusion coefficient assumed to be equal for the two
species.  The initial conditions are that $a(x,t)=const = a_0$ for $x<0$
and
$a(x,t)=0$ for $x\geq0$.  Similarly, $b(x,t)=const = b_0$ for $x>0$ and
$b(x,t)=0$ for $x\leq0$. With these conditions, no matter the
dimensionality of the system, the system of equations is effectively
one-dimensional.

The front problem was first analyzed via a scaling
description~\cite{galfi} and later refined by a large number of
authors using more rigorous theoretical and careful numerical
approaches (see references in Yuste \emph{et al.}~\cite{we2}).
One upshot of the extensive work is that $d=2$ is a critical dimension
for the mean field description to be appropriate. Below $d=2$
one must take into account fluctuations, neglected in this description,
that completely change the outcome of the analysis.
While it has been assumed that
the mean field model holds for the critical dimensions $d=2$,
Krapivsky~\cite{krapivsky}
finds logarithmic corrections that have also been observed in
simulations~\cite{cornell1}.
Our program is to generalize the mean field description to the
subdiffusive case and to apply a scaling description to the problem.
One can show that $d=2$ is also the
critical dimension for the subdiffusive problem.  Our work is therefore
not appropriate for systems of dimension lower than two (we do not
consider logarithmic corrections).

In order to generalize the reaction-diffusion problem to
reaction-subdiffusion, we must deal with the subdiffusive
motion of the particles (generalization of the first term in
Eq.~(\ref{difmft})) and with their reaction rate law (second term).
We discuss each separately.

In the fractional diffusion approach to subdiffusive motion one replaces
the diffusion operator by the Riemann-Liouville operator:
\begin{equation}
D\displaystyle\frac{\partial^2}{\partial x^2} a(x,t) \to
K_\gamma
\displaystyle ~_{0}D_{t}^{1-\gamma }
\frac{\partial^2}{\partial x^2} a(x,t),
\end{equation}
where $K_\gamma$ is the generalized
diffusion coefficient that appears in Eq.~(\ref{meansquaredispl}), and
$~_{0}D_{t}^{1-\gamma } $ is the Riemann-Liouville operator,
\begin{equation}
~_{0}D_{t}^{1-\gamma } f(t)=\frac{1}{\Gamma(\gamma)}
\frac{\partial}{\partial t} \int_0^t d\tau
\frac{f(\tau)}{(t-\tau)^{1-\gamma}}.
\label{LR}
\end{equation}
The reaction-diffusion equations~(\ref{difmft}) are thus replaced by
\begin{equation}
\label{subdifmft}
\begin{array}{rcl}
\displaystyle\frac{\partial}{\partial t} a(x,t)&=&
K_\gamma
\displaystyle ~_{0}D_{t}^{1-\gamma }
\frac{\partial^2}{\partial x^2} a(x,t)-R_\gamma(x,t) \\
\noalign{\smallskip}
\displaystyle\frac{\partial}{\partial t} b(x,t)&=&
K_\gamma
\displaystyle ~_{0}D_{t}^{1-\gamma }
\frac{\partial^2}{\partial x^2} b(x,t)-R_\gamma(x,t) .
\end{array}
\end{equation}
The reaction term $R_\gamma(x,t)$ will be discussed
subsequently
because certain aspects of the problem are independent of the specific
form of this term.

\subsection{Scaling independent of reaction term}
\label{model1}
As the reaction proceeds, a depletion zone develops around the front.
This is the region where the concentrations of reactants are
significantly smaller than
their initial values. How the width $W$ evolves with time is one of
the measures typically used to characterize the process.  Within this
depletion zone lies the so-called reaction zone, the region where the
concentration $c(x,t)$ of the product $C$ is
appreciable.  This concentration profile has a width $w$ whose variation
with time is another characteristic of the evolving reaction. The
evolution of the production rate of $C$ (which determines the height
of the profile of $c(x,t)$ in the reaction zone) is a third measure of
the process.  To find these time dependences
we adapt the original scaling approach~\cite{galfi,cornell1}
to the subdiffusive case, and assume the scaling forms
\begin{equation}
\label{abscaling}
\begin{array}{rcl}
a(x,t)&=& t^{-\theta} \hat{a} ((x-x_f) t^{-\alpha}) \\
\noalign{\smallskip}
b(x,t)&=& t^{-\theta} \hat{b} ((x-x_f) t^{-\alpha})
\end{array}
\end{equation}
for the concentrations and
\begin{equation}
\label{Rgamma}
R_\gamma(x,t)=t^{-\mu} \hat{R}_\gamma ((x-x_f) t^{-\alpha})
\end{equation}
for the reaction term.  The exponents $\theta$, $\alpha$, and $\mu$
are to be determined from three relations.  The scaling forms
are only valid for $x\ll W$,
that is, well within the depletion zone.

Two of the three relations needed to fix the scaling exponents do not
require further specification of the reaction term.
Since the reaction zone increases more slowly than the width of the
depletion zone (an assumption that ex post turns out to be correct), we
can focus on the concentration difference $u(x,t)=a(x,t)-b(x,t)$ to
deduce the width of the latter.  The reaction term drops out when one
subtracts the equations in Eq.~(\ref{subdifmft}), and its form therefore
does not matter at this point.  Generalizing the procedure
of G\'alfi~\cite{galfi}, one can scale
the resulting equation by measuring concentrations in units of $a_0$,
time in units of $\tau=1/(ka_0)$, and length in units of
$l=(K_\gamma\tau^\gamma)^{1/2}$, so that the equation is simply
\begin{equation}
\displaystyle\frac{\partial}{\partial t} u(x,t)=
\displaystyle ~_{0}D_{t}^{1-\gamma }
\frac{\partial^2}{\partial x^2} u(x,t)
\end{equation}
and the only control parameter is $q=b_0/a_0$ in the initial condition:
\begin{equation}
\label{difference}
\begin{array}{rcll}
u(x,0)&=& 1&{\rm for}~~~ x<0 \\
\noalign{\smallskip}
u(x,0)&=& -q &{\rm for}~~~ x>0.
\end{array}
\end{equation}
The solution is
\begin{equation}
u(x,t)= -q +\frac{1+q}{2}
H^{1,0}_{1,1}\left[\frac{x}{t^{\gamma/2} }
        \LM{(1,\frac{\gamma}{2})}{(0,1)}   \right],
\label{ctPeriodica}
\end{equation}
where $H^{1,0}_{1,1}$ is the Fox
H-function~\cite{Sch,MathaiSaxena}.
When $\gamma=1$ this reduces to the diffusion result~\cite{galfi}
\begin{equation}
u(x,t)= -q +\frac{1+q}{2}\;{\rm erfc}\left( \frac{x}{2t^{1/2}}\right).
\end{equation}
From Eq.~(\ref{ctPeriodica}) we see that the width of the depletion
zone scales as
\begin{equation}
W\sim t^{\gamma/2},
\label{width}
\end{equation}
i.e., $\partial a(x,t)/\partial x
\sim \partial b(x,t)/\partial x \sim t^{-\gamma/2}$.
Then, from Eq.~(\ref{abscaling}), the following relation between scaling
exponents follows immediately:
\begin{equation}
\theta + \alpha = \frac{\gamma}{2}.
\label{scaling1}
\end{equation}

The second relation follows from the fact that the concentration
gradient of $A$ and $B$ leads to a flux of particles toward the reaction
region. The assumption that the reaction is fed by these particle
currents
then leads to the quasistationary form in the reaction zone
\begin{equation}
\label{subdifmftstat}
\begin{array}{rcl}
0&=&
K_\gamma
\displaystyle ~_{0}D_{t}^{1-\gamma }
\frac{\partial^2}{\partial x^2} a(x,t)-R_\gamma(x,t) \\
\noalign{\smallskip}
0&=&
K_\gamma
\displaystyle ~_{0}D_{t}^{1-\gamma }
\frac{\partial^2}{\partial x^2} b(x,t)-R_\gamma(x,t) \; ,
\end{array}
\end{equation}
which requires that
\begin{equation}
\mu=\theta+2\alpha+1-\gamma\;.
\label{scaling2}
\end{equation}

For the width of the reaction zone to grow more slowly than the
depletion zone caused by subdiffusion requires that
\begin{equation}
\alpha < \gamma/2\;.
\label{ineq}
\end{equation}
On the other hand, the quasistationarity condition requires that
\begin{equation}
K_\gamma
\displaystyle ~_{0}D_{t}^{1-\gamma }
\frac{\partial^2}{\partial x^2} a(x,t) \sim
t^{-(\theta+2\alpha+1-\gamma)} \gg
\displaystyle\frac{\partial}{\partial t} a(x,t) \sim t^{-(\theta+1)},
\end{equation}
which again leads to Eq.~(\ref{ineq}).

An experimentally accessible quantity that is independent of $R_\gamma$
is the
location $x_f$ of the point at which the production rate of $C$ is
largest.
This should occur where $a(x,t)\sim b(x,t)$, that is, $u(x_f,t)\sim
0$.  The time dependence of this equimolar point is found from
Eq.~(\ref{ctPeriodica}) to be
\begin{equation}
x_f(t)=K_ft^{\gamma/2}
\end{equation}
where $K_f$ is determined from the equation
\begin{equation}
\frac{2q}{1+q}=
H^{10}_{11}\left[K_f \LM{(1,\frac{\gamma}{2})}{(0,1)} \right].
\end{equation}

\subsection{Choice of reaction term and resultant scaling}
\label{model2}

Further relations involving the scaling exponents aimed at their
expression in terms of model quantities require specification of the
reaction term. There is a varied literature on this subject, based on a
number of different assumptions~\cite{wearne,vlad,fedotov,sung,seki1,seki2}.
Most do not associate a memory with the reaction term. Some
assume that, as in the case of ordinary diffusion, reactions can simply
be modeled by a space-dependent form of the law of mass action, e.g.,
by setting $R=ka(x,t)b(x,t)$.  Some of these assumptions may be
appropriate if the reaction is very rapid, but not if many encounters
between reactants are required for the reaction to occur.

We adopt the viewpoint put forth in a recent theory developed for
geminate recombination~\cite{seki1,seki2} but,
as the authors themselves point out, much more broadly applicable.
This theory
goes back to the continuous time random walk picture from which the
fractional diffusion equation can be obtained, and considers
\emph{both} the motion and the reaction in this framework.
In the context of geminate recombination the authors define a reaction
zone and argue that a geminate pair within the reaction zone will not
necessarily react for any finite intrinsic reaction rate
(which they call $\gamma_{rc}$) because one of the
particles may leave the zone before a reaction takes place. The dynamics
of leaving the reaction zone is ruled by the waiting time distribution
$\psi_{out}(t)=\psi(t)e^{-\gamma_{rc}t}$ where $\psi(t)$ is the waiting
time that regulates the rest of the dynamics [cf.
Eq.~(\ref{waitingtime})], and therefore the reaction
rate will acquire a memory that arises from the same source as the
memory associated with the subdiffusive motion.  In the continuum limit
this model then leads to a reaction-subdiffusion equation in which both
contributions have a memory.  Seki \emph{et al.}~\cite{seki1,seki2}
obtain a subdiffusion-reaction
equation which at long times corresponds to choosing a reaction term of
the form
\begin{equation}
R_\gamma(x,t)=k ~_{0}D_{t}^{1-\gamma }a(x,t)b(x,t).
\label{specific}
\end{equation}
Here ``long times'' set in very quickly if the reaction zone is narrow
and the intrinsic reaction rate small.  As noted earlier, although the
derivation is specifically for geminate recombination, the arguments can
be generalized.

Our full reaction-subdiffusion starting equations
on which the remainder of this paper is based then are
\begin{equation}
\label{subdifmftfull}
\begin{array}{rcl}
\displaystyle\frac{\partial}{\partial t} a(x,t)&=&
~_{0}D_{t}^{1-\gamma }
\left\{K_\gamma\displaystyle\frac{\partial^2}{\partial x^2}
a(x,t)-ka(x,t)b(x,t)\right\} \\
\noalign{\smallskip} \noalign{\smallskip}
\displaystyle\frac{\partial}{\partial t} b(x,t)&=&
~_{0}D_{t}^{1-\gamma }
\left\{K_\gamma\displaystyle\frac{\partial^2}{\partial x^2}
b(x,t)- -ka(x,t)b(x,t)\right\}\; .
\end{array}
\end{equation}
From the specific reaction term given in Eq.~(\ref{subdifmftfull}) we
can now obtain the third relation
between the scaling exponents by balancing the terms within the
brackets:
\begin{equation}
\mu = 2\theta +1-\gamma.
\label{scaling3}
\end{equation}
Simultaneous solution of Eqs.~(\ref{scaling1}), (\ref{scaling2}), and
(\ref{scaling3}) finally yields
\begin{equation}
\alpha=\frac{\gamma}{6}, \qquad \theta=\frac{\gamma}{3}, \qquad
\mu=1-\frac{\gamma}{3}.
\label{thescales}
\end{equation}

\section{Comparisons with numerical solutions}
\label{comparisons}
Elsewhere we have carried out exhaustive numerical simulations of the
reaction-subdiffusion system and compared our scaling results to those
of the simulations for a variety of front properties~\cite{we2}.
The simulations
were described in detail and are particularly subtle when motions are
slow and concentrations are low.  The agreement with our scaling
results
was in some cases quantitative and in others, particularly where very
small numbers are involved, at least qualitative.

\begin{figure}
\begin{center}
\includegraphics[width=0.6\columnwidth]{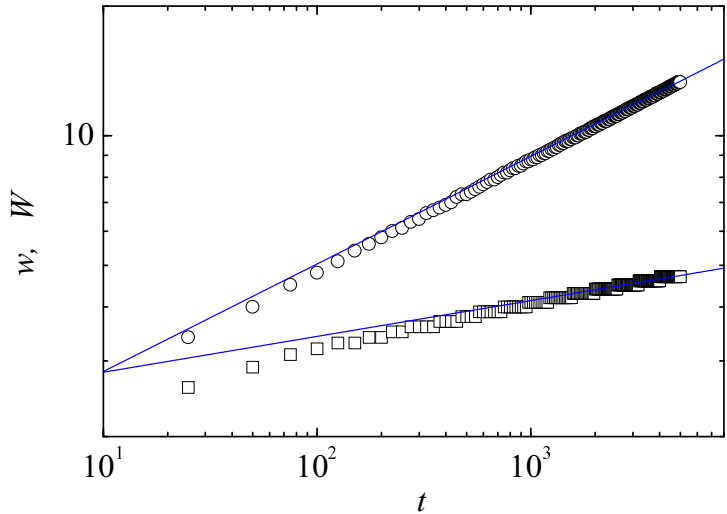}
\end{center}
\caption{Log-log plots of the width of the product profile $w(t)$ and
the width of the depletion zone $W(t)$ vs $t$.  The squares are the
numerical results for $w(t)$ and the comparison line has slope $\gamma/6$.
The circles are the numerical results for $W(t)$ and the line has
slope $\gamma/2$.
\label{wWga05}}
\end{figure}

\begin{figure}
\begin{center}
\includegraphics[width=0.6\columnwidth]{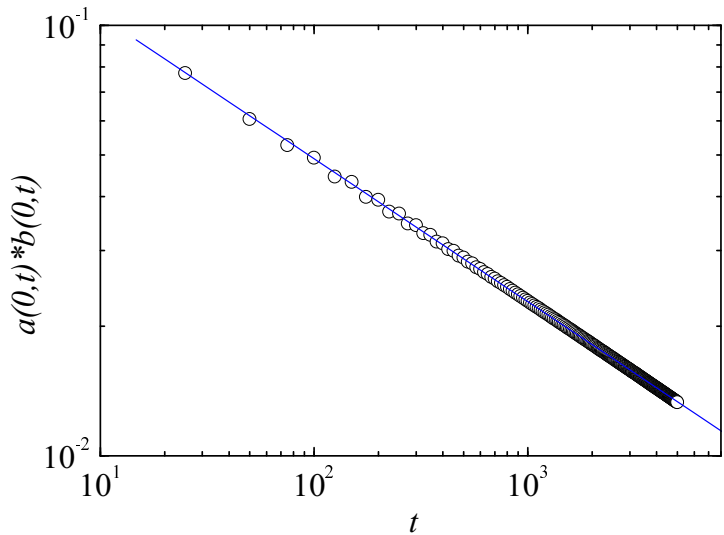}
\end{center}
\caption{Log-log plots of the product $a(0,t)\times b(0,t)$ vs $t$.
The circles are the numerical results and the line has the slope
$2\gamma/3$ predicted for this product by our scaling theory.
\label{a0tb0t}}
\end{figure}

Here we present a different comparison, namely, between our
exact results (where we have them) and our scaling
results, and those obtained by direct numerical integration of the
reaction-subdiffusion model~\cite{numerical}.  This analysis
provides an insight not only
into the accuracy of the assumptions that enter the scaling analysis,
but also into the length of time (rather short, as it turns out) before
the asymptotic scaling results are valid.  All of our numerical
results here have been obtained for the exponent $\gamma=0.5$ and with
the parameters $K_{1/2}=k=1$.

Figure~\ref{wWga05} shows the width of the product profile, $w(t)$,
predicted to grow as $t^{\gamma/6}$, and the width of the depletion
zone, $W(t)$, which should grow as $t^{\gamma/2}$ according to our
scaling theory.  The asymptotic prediction for the former is
very good after about $t=100$ and for the latter after about $t=200$.
Note that whereas the scaling prediction for $W(t)$ is independent of
the form of the reaction term, that of the product profile width $w(t)$
is not.

The product $a(0,t)\times b(0,t)$ is shown in Fig.~\ref{a0tb0t}.  The
agreement between the numerical integration and the scaling result is
excellent from the earliest times indicated.

\begin{figure}
\begin{center}
\includegraphics[width=0.6\columnwidth]{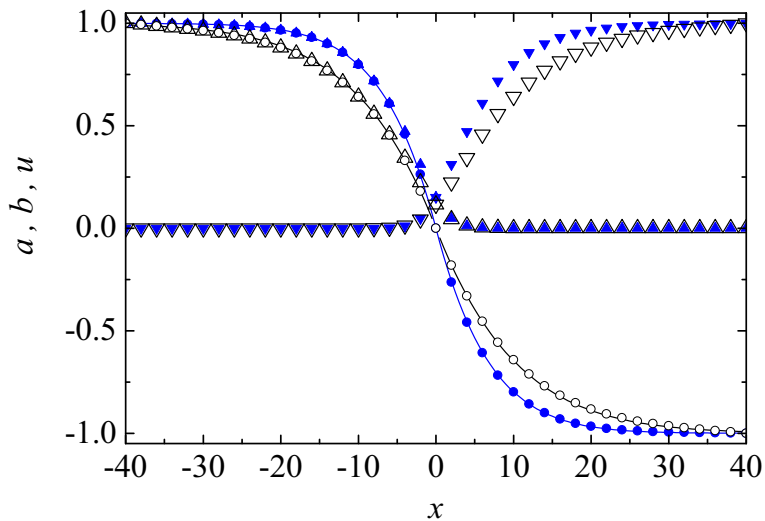}
\end{center}
\caption{Profiles $a(x,t)$, $b(x,t)$, and the difference profile
$u(x,t)=a(x,t)-b(x,t)$ vs $x$ for two different times.  The detailed
description of this figure is in the text.  The theoretical
vs numerical comparison is between the circles and the solid curves.
\label{abuga05}}
\end{figure}

Figure~\ref{abuga05} contains a more elaborate set of results
showing the space dependence of various quantities at two different
times.  The triangles pointing upward represent the reactant
concentration $a(x,t)$ as a function of position $x$. As expected, this
profile is large on the left side of the graph (since reactant $A$ was
originally located to the left of the front).  The upper curve (solid
triangles)
corresponds to time $t=1000$ and the lower curve (white triangles),
when more of $A$
has already reacted, corresponds to $t=5000$.  On the other side and in
complete mirror image (because we took both initial concentrations to be
equal) are the corresponding results for reactant $B$. The triangles
pointing downward represent $b(x,t)$ at the two times, with the same
time progression from more reactant to less reactant as time moves on.
The quantity that we can directly
compare to our theory is the difference profile
$u(x,t)=a(x,t)-b(x,t)$. The numerical integration results are represented
by the circles. Way to the left of the diagram $u(x,t)\approx a(x,t)$
and way to the right $u(x,t)\approx b(x,t)$, but near the origin both
concentrations contribute to $u(x,t)$.  The outer solid curve
(through solid circles) is the
exact result Eq.~(\ref{ctPeriodica}) at $t=1000$ and the inner solid
curve is for $t=5000$. The agreement is perfect, as it should be.

\begin{figure}
\begin{center}
\includegraphics[width=0.6\columnwidth]{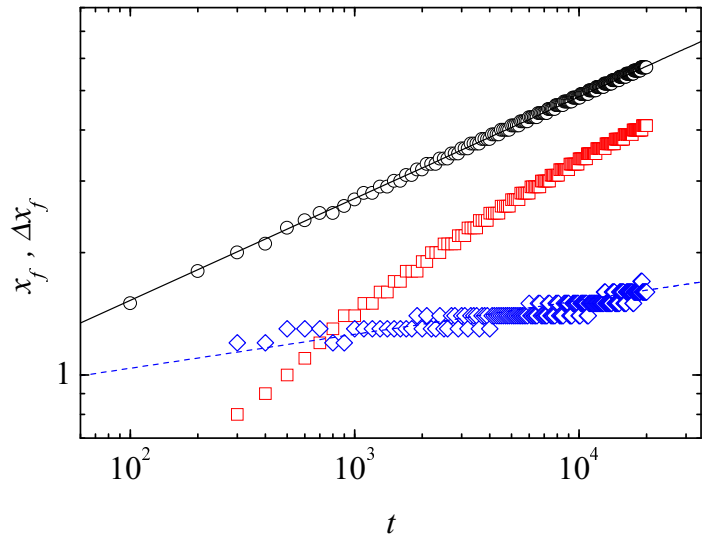}
\end{center}
\caption{Equimolar point $x_f(t)$ vs $t$: circles are numerical
integration results and the solid line our exact theoretical values.
The squares are the numerical results for the point of maximum product
formation, $\hat{x}_f(t)$.  The difference between the two is represented
by the diamonds, and the dashed line is our scaling prediction of slope
$\gamma/6$.
\label{xfga05}}
\end{figure}

Finally, in Fig.~\ref{xfga05} we present results for the time dependence
of two quantities frequently used to characterize the point of maximum
rate of product formation.  The point one often sees chosen for this
characterization is $x=x_f$ such that $a(x_f,t)=b(x_f,t)$.  This
time-dependent equimolar point can be found exactly from the solution
Eq.~(\ref{ctPeriodica}), $x_f=K_ft^{\gamma/2}$, where $K_f$ is determined
from the equation
\begin{equation}
\frac{2q}{1+q}=
H^{10}_{11}\left[ K_f \LM{(1,\frac{\gamma}{2})}{(0,1)} \right].
\end{equation}
The circles in the figure are the numerical results for $x_f(t)$ with
$q=1/2$ and the
line is our theoretical prediction with $K_f=0.482742\ldots$.  A second
point (and one that is actually a more accurate characterization of the
maximum rate of product formation) is $\hat{x}_f$ such that the product
$a(\hat{x}_f,t)b(\hat{x}_f,t)$ which determines the reaction
rate is a maximum.  This
quantity is more difficult to predict, but we do know that both $x_f$
and $\hat{x}_f$ are within the reaction zone, which scales as
$x/t^{\gamma/6}$. It therefore follows that the distance between these
two points should grow as $x_f - \hat{x}_f \sim t^{\gamma/6}$.  But
since $x_f\sim t^{\gamma/2}$, we conclude that $\hat{x}_f \approx x_f$
only for times for which their difference is negligible, i.e. only
for times such that $t^{-2\gamma/3} \gg 1$.  In Fig.~\ref{xfga05} the
squares are the numerical results for $\hat{x}_f(t)$, the diamonds
represent the difference $x_f - \hat{x}_f$, and the dashed line has
slope $\gamma/6$. For the two measures to approach closely one has to go
to considerably longer times.

\section{End notes}
\label{endnotes}
We have presented a scaling theory for the evolution of a
reaction-subdiffusion front for the annihilation reaction $A+B\to C$.
The theory is based on a set of fractional equations in which the
diffusion operator is replaced by a Riemann-Liouville operator, and the
reaction term is adjusted as well to take into account the slower motion
of the reactants.  Elsewhere we presented a comparison of our
theoretical results with numerical simulations of the process~\cite{we2}.
Here we
have presented a detailed comparison of our results with those of
numerical integrations of the fractional equations~\cite{numerical}.
The familiar rate
anomalies of the reaction-diffusion problem are now more
pronounced by the slower motion of the reactants, with the result that
scaling behaviors are recovered from those of the corresponding
diffusive problem with the substitution $t\to t^\gamma$.

\begin{acknowledgments}
This work was partially supported by the Engineering Research Program of
the Office of Basic Energy Sciences at the U. S. Department of Energy
under Grant No. DE-FG03-86ER13606, and by the Ministerio de Ciencia y
Tecnolog\'{\i}a (Spain) through Grant No. FIS2004-01399.
\end{acknowledgments}

\end{document}